\documentclass[aps,superscriptaddress,amsmath,amssymb,twocolumn,showpacs,floatfix,reprint]{revtex4-2}

\usepackage{comment}

\usepackage{bm}
\usepackage{float}
\usepackage{graphicx}
\usepackage[colorlinks=true, urlcolor=blue, linkcolor=blue, citecolor=blue, pdftex]{hyperref}
\usepackage{cleveref}  
\usepackage{nicefrac}
\usepackage{multirow}
\usepackage{physics}
\usepackage[usenames,dvipsnames]{xcolor}

\definecolor{C0}{HTML}{1f77b4}
\definecolor{C1}{HTML}{ff7f0e}
\definecolor{C2}{HTML}{2ca02c}
\definecolor{C3}{HTML}{d62728}
\definecolor{C4}{HTML}{9467bd}
\definecolor{C5}{HTML}{8c564b}
\definecolor{C6}{HTML}{e377c2}
\definecolor{C7}{HTML}{7f7f7f}
\definecolor{C8}{HTML}{bcbd22}
\definecolor{C9}{HTML}{17becf}
\definecolor{gaussian}{HTML}{47908C}

\newcommand{\EE}{\mathbb{E}\,}
\newcommand{\reals}{\mathbb{R}}

\def\be{{\mathbf{e}}}

\def\bh{{\mathbf{h}}}

\def\bp{{\mathbf{p}}}

\def\bs{{\mathbf{s}}}
\def\bt{{\mathbf{t}}}

\def\bv{{\mathbf{v}}}

\def\bx{{\mathbf{x}}}

\begin{document}

\title{Mapping of attention mechanisms to a generalized Potts model}

\author{Riccardo Rende}
\author{Federica Gerace}
\author{Alessandro Laio}
\author{Sebastian Goldt}
\email{sgoldt@sissa.it}
\affiliation{Scuola Internazionale Superiore di Studi
  Avanzati (SISSA), Via Bonomea 265, 34136 Trieste, Italy}

\date{\today}


\begin{abstract}
  Transformers are neural networks that revolutionized natural
  language processing and machine learning. They process sequences of
  inputs, like words, using a mechanism called self-attention, which is trained
  via masked language modeling (MLM). In MLM, a word is randomly masked in an
  input sequence, and the network is trained to predict the missing word.
  Despite the practical success of transformers, it remains unclear what type of
  data distribution self-attention can learn efficiently. 
  Here, we show
  analytically that if one decouples the treatment of word positions and embeddings, a single layer of self-attention
  learns the conditionals of a generalized Potts model with interactions between
  sites and Potts colors. Moreover,  we show that training this neural network  is exactly equivalent to solving the inverse Potts problem by   the so-called pseudo-likelihood method, well known in statistical physics. Using this mapping, we 
  compute the generalization error of self-attention in a model scenario analytically using the replica method.
\end{abstract}

\maketitle

\begin{figure}[!b]
  \vspace*{-1em}
  \includegraphics[width=\columnwidth, trim = 0cm 3cm 0cm 0cm, clip]{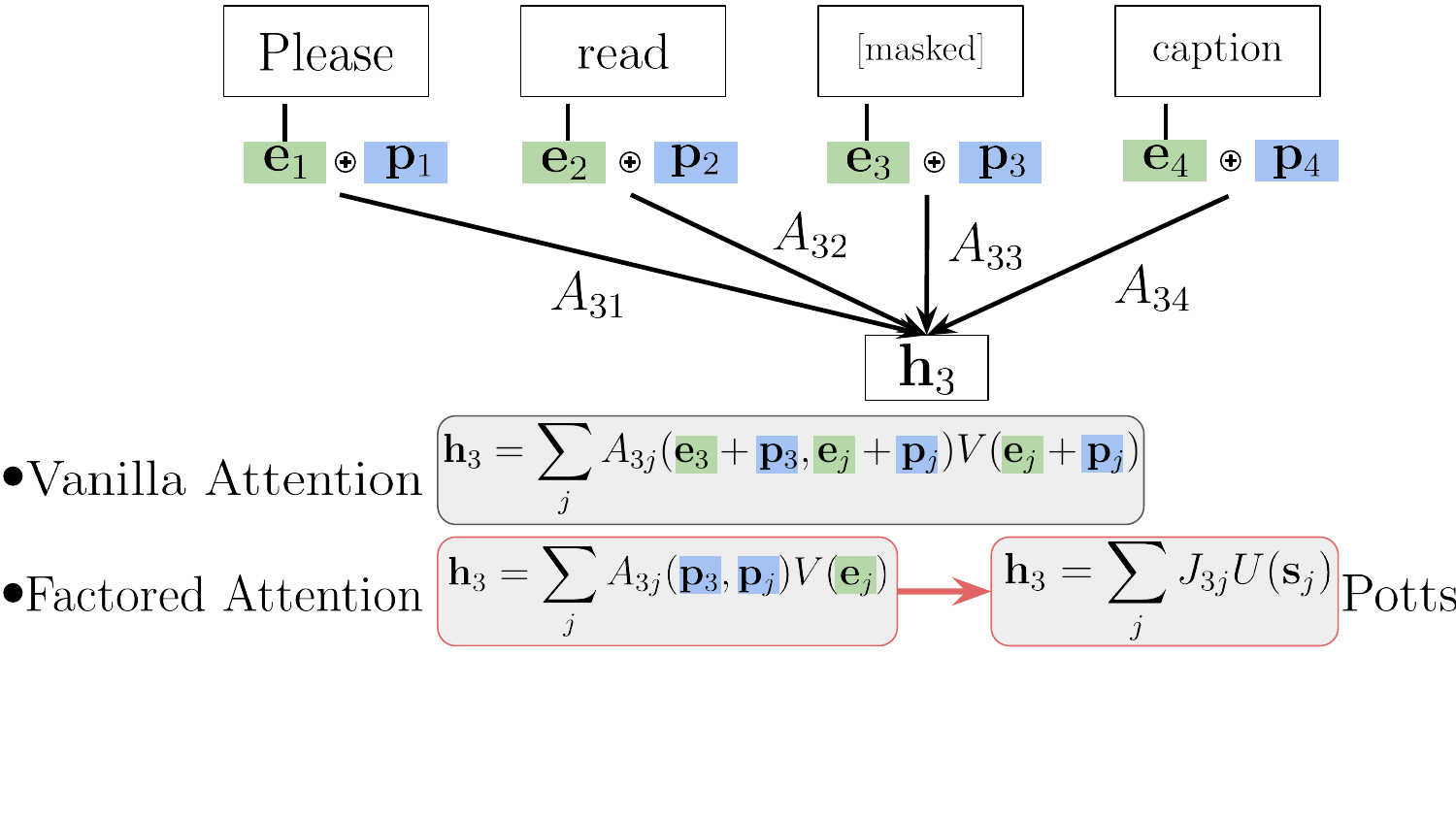}
  \vspace*{-1em}
  \caption{\label{fig:figure1} \textbf{Masked language modeling (MLM) with a
      single layer of self-attention.} The goal of MLM is to predict the masked
    word in a given sentence. Self-attention first maps words into
    representations $\be_j + \bp_j$, where $\be_j$ are embedding
    vectors representing words, and $\bp_j$ encode their positions. For a
    given masked word, the associated attention vector $\bh_{k}$ is computed as a linear combination of the values $\bv_j = V(\be_j + \bp_j)$ of all other tokens, weighted by the attention weights $A_{kj}$. In vanilla self-attention, values and attention weights depend on embeddings \emph{and} positional vectors, while in {\it factored} attention, attention weights depend only on positions, and values only on the embeddings. By identifying the attention weights~$A$ with the interaction matrix $J$ of a Potts model~\cref{eq:potts}, the value matrix $V$ with the color similarity matrix U and the embedding vectors with the one-hot spins, we get a learning model identical to a Potts model.}
\end{figure}

\paragraph*{Introduction.} Transformers~\cite{vaswani2017attention} are a
powerful type of neural network that have achieved state-of-the art results in
natural language processing (NLP)~\cite{devlin2018bert, howard2018universal,
  radford2018improving, brown2020language, chatgpt}, image
classification~\cite{dosovitskiy2021an}, and even protein structure
prediction~\cite{jumper2021highly}.  While standard neural networks can be
thought of as functions of a single input, transformers act on sets of
``tokens'', like words in a sentence. The key to the success of transformers is
a technique called masked language modeling (MLM), where transformers are
trained to predict missing words in a sentence~\cite{devlin2018bert,
  howard2018universal, radford2018improving, brown2020language, chatgpt}, cf.\
\cref{fig:figure1}a. This technique has the advantage that it can leverage large
amounts of raw text (or images, or protein sequences) without any annotation. By learning the conditional
distribution of having a word in a specific position of the sentence, given the
other words, transformers ostensibly learn the relationships between words in a
robust way.

The basic building block of transformers is the \emph{self-attention} (SA)
mechanism~\cite{bahdanau2014neural, kim2017structured}, which transforms a sequence
of tokens $\bx_j$ into another sequence $\bh_j$. We illustrate self-attention on a masked language
modeling task in \cref{fig:figure1}. The sentence is first transformed into a
set of representations $\bx_j = \be_j + \bp_j$, where $\be_j$ is a
vector representing the $j$th word and the vector $\bp_j$ encodes its
position. SA then computes a linear transformation of the
representations to yield the values~$\bv_j$. The $k$th output vector $\bh_k$ is then a
linear combination of the values~$\bv_j$ weighted by an attention matrix~$A$,
whose elements $A_{kj}$ quantify the relative importance of the $j$th
input token for the $k$th output vector, for example based on their semantic
similarity. The functions to compute values~$\bv_j$ and the attention
matrix~$A$ both have trainable parameters; see \cref{eq:sa} for a precise
definition. The flexibility of self-attention comes from the attention weights~$A_{kj}$, which are not fixed, but computed given the  context, i.e.\ the surrounding tokens.

The practical success of transformers raises several fundamental questions: what
are the statistical structures that self-attention learns with MLM? More
precisely, since the MLM objective is to learn the conditional probability
distribution of words given a set of surrounding words, which family of conditional probabilities can self-attention learn? And how many samples are required to achieve good performance? Here, we make a step towards answering these questions by exploiting tools from the statistical physics of learning~\cite{gardner1989three, seung1992statistical,
  engel2001statistical, carleo2019machine}.

The first challenge is to design a data model that mimics the structure of real
sentences. While classical works modelled inputs as vectors of i.i.d.\ random
variables, recent work has introduced more sophisticated data models for
neural networks~\cite{chung2018classification, goldt2020modeling,
  spigler2020asymptotic, gerace2020generalisation,
  favero2021locality, loureiro2021learning,
  ingrosso2022data} which allowed the study of unsupervised
learning~\cite{refinetti2022dynamics, cui2023high}. To analyze the
self-supervised learning of MLM, we model sequences of words as system of spins, interacting via a generalized Potts Hamiltonian~\cite{potts1952some, wu1982potts} with couplings between colors
(=words) and positions. We sample a synthetic data set from the Potts model
using Monte Carlo, and we perform masked language modeling by training a
transformer to predict masked spins in spin sequences. While an off-the-shelf transformer
requires several layers of self-attention to learn this simple probability
distribution, we show analytically that a single layer of \emph{factored} self-attention, where
we separate the treatment of positions and inputs, can reconstruct the couplings of the Potts model
\emph{exactly} in the limit of a large training set. In particular, we derive an
exact mapping between the output of the self-attention mechanism and the
conditional distribution of a  Potts spin given the others. We finally use this mapping to compute the generalization loss of a
single layer of self-attention analytically using the replica method.

\paragraph*{A generalized Potts model to sample sequences.} We model sentences
as sequences of spins $\mathbf{s}=(\mathbf{s}_1, \dots, \mathbf{s}_L)$, with
$\mathbf{s}_i\in\reals^C$ taking values from a vocabulary of~$C$ colors, which
we encode as one-hot vectors. Each color can be thought of as a word in natural
text, an amino acid in a protein, etc. In a standard Potts Hamiltonian, only
spins of the same color interact with each other via an interaction matrix
$J$. This is an unrealistic model for real data: it treats all colors as
orthogonal, even though words and amino acids have varying degrees of
similarity. We therefore generalize the Potts Hamiltonian to
\begin{equation}
  \label{eq:potts}
  \mathcal{H}\left( \mathbf{s} \right) = - \frac{1}{2} \sum_{i,j=1}^L J_{ij} \mathbf{s}^T_{i} U \mathbf{s}_{j},
\end{equation}
where $J \in \reals^{L\times L}$ governs the interactions between spins
at different positions, and $U \in \reals^{C\times C}$ encodes the
similarities between colors (we denote matrices by capital letters and vectors
in boldface). Without loss of generality, we set $J_{ii}=0$ and sample
sequences from the Boltzmann distribution
$P\left( \mathbf{s} \right) \propto \exp\left[ - \beta
  \mathcal{H}(\mathbf{s})\right]$.  We recover the standard Potts model by
choosing $U$ as the identity matrix.

\paragraph*{Masked language modeling with transformers.} Given the generative
model~\eqref{eq:potts}, the MLM objective amounts to predicting the $i$th spin
given the sequence $\mathbf{s}_{\setminus i}$ where that spin is
``masked'', i.e.\ $\bs_i = \bt$, the masking token.
To apply self-attention to a sequence $\bs_{\setminus i}$, we first compute the
values $\bv_j = V (E\mathbf{s}_{j} + a\mathbf{p}_j)$, where the embedding matrix~$E \in \reals^{d \times C}$ maps Potts colors into $d$-dimensional representation vectors, and $V \in \reals^{d \times d}$ is a weight matrix; both $E$ and $V$ are
trainable parameters. The scalar parameter $a$ controls the relative importance between the embedding and positional encoding vectors. The output vector $\bh_i$ corresponding to the masked token is a linear
combination of the values, weighted by an exponential attention function~\cite{vaswani2017attention}:
\begin{equation}
  \label{eq:sa}
  \bh_i\left(\mathbf{s}_{\setminus i}\right) =  \sum_{j=1}^{L}\frac{\exp\left[\left(E\mathbf{t}+\mathbf{p}_{i}\right)^\top Q^\top K \left(aE\mathbf{s}_{j}+\mathbf{p}_{j}\right)\right]}{\sum_{k}\exp\left[\left(E\mathbf{t}+\mathbf{p}_{i}\right)^\top Q^\top K \left(aE\mathbf{s}_{k}+\mathbf{p}_{k}\right)\right]} \bv_j.
\end{equation}
Crucially, the $i$th spin $\bs_i$ in this expression has to be replaced with the masking token~$\mathbf{t}$, since it is the masked input. The matrices~$Q, K \in \reals^{d \times d}$
are also trainable parameters of the model. In the following we take the embedding dimension equal to the number of colors, $d=C$, in order to be able to map the output vector $\bh_i$ into a probability distribution~$\tilde \bp_i$ over the colors through the \emph{softmax} non-linearity~\footnote{For a vector $\bx=(x_i)$, the softmax non-linearity yields $x'_i=\exp(x_i) / \sum_j \exp(x_j)$. The elements of $\bx'$ can be interpreted as a probability distribution, since they are positive and sum to one.}.

\begin{figure*}[ht!]
  \includegraphics[width=\linewidth]{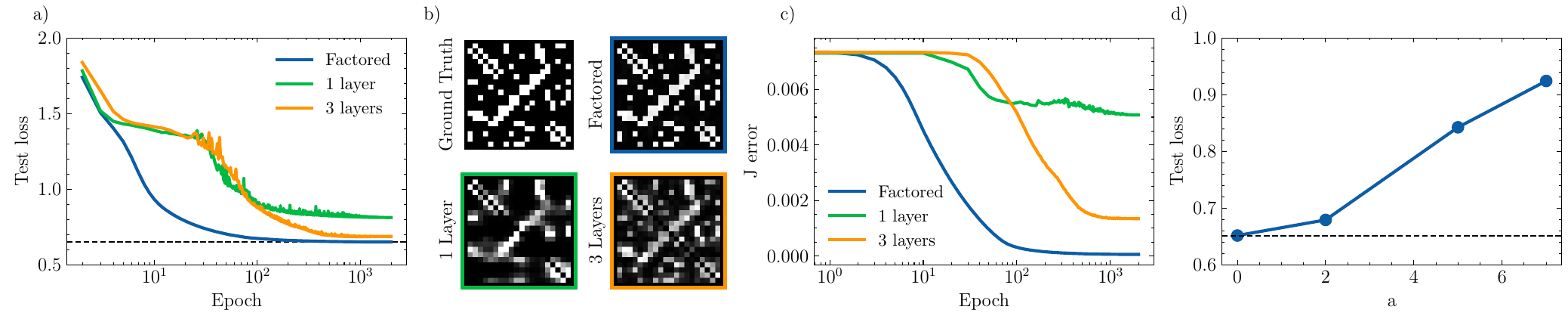}
  \caption{\label{fig:factored_vs_vanilla} \textbf{A single layer of factored
      self-attention learns the generalized Potts model efficiently.}
    \textbf{(a)} Test loss~\eqref{eq:eg} for factored self-attention and for
    vanilla transformers with one and three layers during training with
    stochastic gradient descent. The optimal generalization loss is shown as a
    black dashed line. \textbf{(b)} Interaction matrix $J$ of the generative
    Potts Model~\eqref{eq:potts} compared to the attention maps learned by
    transformers with vanilla and factored self-attention. For the three-layer
    transformer, the attention map was obtained by averaging the maps of the
    last two layers. \textbf{(c)} Reconstruction error of the interaction
    ${(J - A)}^2$ as a function of the number of epochs for all considered
    architectures. \textbf{(d)} Test loss as a function of perturbation level
    $a$. Decoupling the treatment between positions and colors by decreasing
    $a$ decreases the test loss. \emph{Parameters:} sequence length $L=20$, vocabulary size $C=20$, embedding dimension $d=20$, $M=3000$ data points.}
  \vspace*{-1em}
\end{figure*}

\paragraph*{Training a vanilla transformer on the generalized Potts model.} For
our first experiment, we emulate the setting of protein structure prediction, so
we choose a vocabulary of size $C=20$ and sample a symmetric interaction matrix
$J_{ij}=\{0, 1\}$ which we show in \cref{fig:factored_vs_vanilla}b). We draw
the entries of the symmetric interaction matrix $U$ i.i.d.\ from the
standard Gaussian distribution. Given these parameters, we use Gibbs sampling to
generate a data
set 
with~$M=3000$ sequences of length $L=20$. We tune the inverse temperature
$\beta$ to ensure an average Hamming distance of~0.3 between sampled sequences,
typical for protein families~\cite{finn2014pfam}.

We then train off-the-shelf transformers consisting of one and three layers on this data set by minimising
the cross-entropy loss between the output distribution and the missing spin
using stochastic gradient descent (see supplementary material for the numerical details)  on the loss
$\mathcal{L}(\mathbf{s}) = - L^{-1} \sum_{i=1}^L\sum_{\alpha=1}^C
s_{i\alpha} \log \tilde{p}_{i\alpha}(\mathbf{s})$, for a sequence
$\mathbf{s}$.  In \cref{fig:factored_vs_vanilla}, we show the test loss
\begin{equation}
  \label{eq:eg}
  \epsilon_g = \mathbb{E}_{\mathbf{s}\sim P} \left[ \mathcal{L}(\mathbf{s})
\right]
\end{equation}
during training, where $\EE_{\mathbf{s} \sim P}$ denotes an average over the
generative model~\eqref{eq:potts}. A transformer with a single layer
of self-attention does not attain the optimal generalization error (black dashed
line). By plotting the attention matrix of the single layer, we
see that the transformer recovers the original interaction matrix to some
degree, albeit not perfectly. Training transformers with three layers on
the same data set improves the accuracy at the cost of loosing interpretability:
there is no straightforward way to collapse several layers of non-linear
transformations of the input sequence into a single attention map; we show the average
of the final two attention layers in \cref{fig:factored_vs_vanilla}b.

\paragraph*{Factored self-attention learns the generalized Potts model.} 
We now consider a variant of self-attention in which the treatment of positions and values is decoupled.
We set $a=0$ in \cref{eq:sa} , set the masking token~$\mathbf{t}=0$, choose one-hot encodings for the positions, and fix the embedding matrix at $E=I_{C}$, so that 
\begin{equation}
  \label{eq:factored_attention}
    h_{i\alpha}\left(\mathbf{s}_{\setminus i}\right) = \sum_{j=1}^L A_{ij} (V \mathbf{s}_{j})_\alpha
\end{equation}
where $A_{ij} \equiv e^{(Q^\top K)_{ij}} / \sum_{k=1}^L e^{(Q\top
  K)_{ik}}$.
This modified self-attention has exactly the same form as the conditional distribution of the generalized Potts model if one sets $U=V$ and~$\beta J~=~A$, which is
 \begin{equation}
   \label{eq:exact_marginals}
     p(s_{i\alpha} = 1 | \mathbf{s}_{\setminus i})= \frac{ \exp\left({
        \beta \sum_{j=1}^L J_{ij} (U \mathbf{s}_{j})_\alpha}\right) }{
       \sum_{\gamma=1}^C \exp \left({\beta \sum_{j=1}^L J_{ij} (U \mathbf{s}_{j})_\gamma }\right) }.
\end{equation}
This equivalence between factored self-attention and the Potts model is our first main result; we now discuss its ramifications.

Decoupling positions and colors 
leads to a significant improvement in the performance of a single layer, which
reaches the optimal generalization error and converges faster, cf.\
\cref{fig:factored_vs_vanilla}a. Factored self-attention recovers the interaction matrix $J$ perfectly, cf. \cref{fig:factored_vs_vanilla}b for the attention map and \cref{fig:factored_vs_vanilla}c for the reconstruction error of the interaction matrix. In
\cref{fig:factored_vs_vanilla}d, we show that decoupling the treatment of positions
and colors completely performs better than any intermediate solution with $a>0$.

Factored attention layers, and thus input-independent attention weights, have been already used as a building block for deep transformers, outperforming standard attention in different applications: \citet{bhattacharya2020single} used it to analyze protein sequences and found that a single layer of factored
self-attention performed as well as a deep transformer, and
significantly better than a single layer of vanilla
self-attention, without explicitly explaining this observation. Moreover, using factored attention is key to obtaining
state-of-the-art results in approximating ground states of many-body quantum
systems~\cite{viteritti2022transformer, rende2023simple, viteritti2023transformer}.

Intriguingly, the form of the loss for masked language modeling with factored self-attention as described above exactly matches the loss of
the pseudo-likelihood method, which has been used for solving the inverse Ising
problem~\cite{besag1975statistical, cocco2011adaptive, ricci2012bethe,
  ekeberg2013improved, cocco2018inverse}.
The pseudo-likelihood method is statistically
consistent~\cite{hyvarinen2006consistency, ravikumar2010high,
  aurell2012inverse}, i.e.\ its parameter estimates converge
to the true parameters as the number of samples goes to infinity. A direct consequence of the mapping in \cref{eq:factored_attention} is thus that
MLM with factored self-attention enjoys the same asymptotic optimality.

\begin{figure*}[!t]
  \includegraphics[width=.85\linewidth]{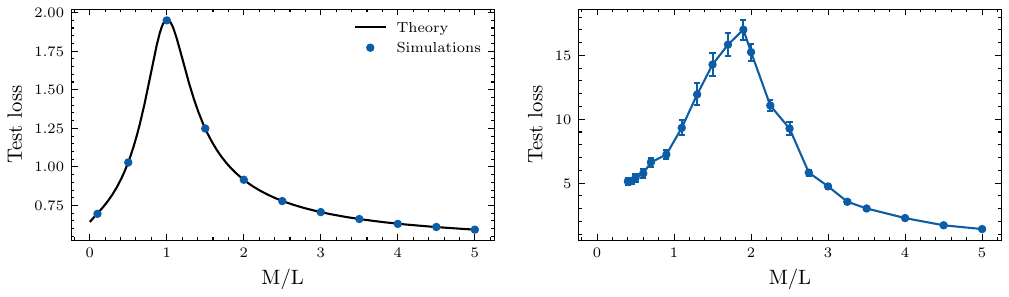}
  \vspace*{-1em}
  \caption{\label{fig:peak_factored} \textbf{The interpolation peak of factored
      attention in theory and practice.} \textbf{(Left)} A replica analysis
    predicts the test loss exactly. Test loss of a single layer of factored
    self-attention as a function of the number of samples per input dimension,
    as computed using replica theory (solid line). The blue points represent the
    outcome of numerical minimisation of the square
    loss~\eqref{eq:simplified_optimization}, averaged over 30 realisations, and
    show perfect agreement with the theory. Error bars are smaller than point
    size. \textbf{(Right)} Same plot for a single layer of factored
    self-attention in the setting of \cref{fig:factored_vs_vanilla} ($L=C=20$),
    showing the same qualitative behaviour. The simulations are averaged over
    $n=30$ different realisations.}
  \vspace*{-1em}
\end{figure*}

\paragraph*{The sample complexity of self-attention.}
A key quantity in machine learning problems is the sample complexity, namely how
many samples are required to achieve a small generalization loss~$\epsilon_g$ with a given model. The mapping 
introduced in this work
allows us to address this question precisely for a single layer of self-attention by means of the replica method from statistical physics.
The
main difficulty in the calculation lies in handling the non-trivial data
distribution~\eqref{eq:potts}. This difficulty can be mitigated thanks to recent advances in statistical physics, which allow us to extend the replica method of disordered systems to structured data \cite{gerace2020generalisation, goldt2022gaussian, loureiro2021learning}. 
To perform the replica calculation, we first relax the discrete nature of Potts spins by re-writing the generalized
Potts Hamiltonian (Eq.~\ref{eq:potts}) in terms of spin magnetization
$\mathbf{m} = \langle \mathbf{s} \rangle_{P\left( \mathbf{s} \right)} $
following mean-field theory. The associated Boltzmann measure then turns into a
multivariate Gaussian distribution, whose covariance matrix is the negative
inverse of the interaction matrix, i.e.\
$\Sigma = - J^{-1}$~\cite{morcos2011direct,
  nguyen2017inverse}. We then draw sequences
$\{\mathbf{m}^{\mu}\}_{\mu = 1}^M$ of length~$L$ from the multivariate Gaussian
with zero mean and covariance matrix
$\Sigma = \left(\Omega L^{-1/2} + \nu \mathbb{I}\right)^{-1}$,
where~$\Omega$ is a symmetric full-rank random matrix sampled from the
Gaussian Orthogonal ensemble while $\nu \mathbb{I}$ is a diagonal matrix
centering the spectrum of $\Omega$ in $\nu$. To ensure $\Sigma$
to be positive definite, we set $\nu > 2$ due to the semicircle
law~\cite{livan2018introduction}. By fixing the location $i$ of the masked spin
across all input sequences, solving the MLM task is equivalent to inferring the
$i$th row of the interaction matrix $\mathbf{J}_i$.

To accomplish this task, we train a single layer of factored self-attention by empirical risk minimisation of a square loss with $\ell_2$-regularisation:
\begin{equation}
    \label{eq:simplified_optimization}
    \hat{\mathbf{A}}_i = \underset{\mathbf{A}_i}{\mbox{argmin}} \left[\frac{1}{2} \sum_{\mu = 1}^M \left( m^{\mu}_i - \mathbf{A}_i \cdot \mathbf{m}^{\mu}_{\setminus i} \right)^2 + \frac{\lambda}{2} \vert\vert \mathbf{A}_i \vert\vert_2^2 \right].
\end{equation}

Our goal is to characterise the generalization
loss~$\epsilon_g$~\eqref{eq:eg} of factored attention with the parameters obtained from minimising the loss \cref{eq:simplified_optimization}. In the high-dimensional limit, where the number of samples and the sequence length $M,L$ tend to infinity while their ratio ~$\alpha \equiv M/L \sim O\left( 1\right)$, we can express~$\epsilon_g$ using
replica theory as a function of four scalar quantities:
\begin{equation}
    \label{eq: generalization_loss_replica}
    \epsilon_g = \rho + q^{\star} - 2r^{\star} + 1/\nu.
\end{equation}
Here, $\nu$ is the center of the spectrum of $\Omega$, $\rho = \mbox{tr}\,\Sigma_{\setminus i}/\left(\nu^2L\right)$ is a
function of the covariance matrix $\Sigma_{\setminus i}$ where we have
removed the $i$th row and column, while $q^{\star}$ and $r^{\star}$ are the
so-called overlap parameters. They correspond to practically measurable quantities over different realisation of the training set, involving the estimator of the $i$th row of the interaction matrix:  
 \begin{equation}
 \begin{split}
     q^{\star} = \frac{\hat{\mathbf{A}}^t_i \Sigma_{\setminus i} \hat{\mathbf{A}}_i}{L}, \qquad  r^{\star} = - \frac{\left(\mathbf{A}^{\star}_i\right)^t \Sigma_{\setminus i} \hat{\mathbf{A}}_i}{\nu L}.
 \end{split}
 \end{equation}
The parameter $r^*$ can be thus interpreted as an overlap between the estimated attention $\hat{\mathbf{A}}_i$ and the ground-truth value  $\mathbf{A}_i$ while $q^*$ is the overlap of the estimated attention, both mediated by the modified covariance matrix $\Sigma_{\setminus i}$.

As we show in the supplementary material, the
values of these order parameters for a given training set of size~$\alpha$ can be
obtained by solving the optimisation problem
\begin{equation}
  \label{eq:free_energy_extremization}
    f_{\beta} = \underset{q,r,\delta q, \hat{q}, \hat{r}, \delta\hat{q}}{\mbox{extr}} \left[
      -\frac{1}{2}\left(\hat{q}\delta q -  q\delta\hat{q}\right) + \nu r\hat{r} + \underset{L\rightarrow
        \infty}{\mbox{lim}} \frac{1}{L} \Psi_{s}  + \alpha \Psi_{e} \right] 
\end{equation}
which yields the \emph{typical} value (over the data) of the free energy density
associated with a Gibbs measure at inverse temperature $\beta$ whose Hamiltonian
corresponds to the loss function in \cref{eq:simplified_optimization}. Note that
the optimisation only involves the scalar parameters $q$, $r$, $\delta q$ and their conjugates~
$\hat{q}$,~$\hat{r}$~and $\delta \hat{q}$, with $\delta q = q - q_s$, $\delta \hat{q} = \hat{q} + \hat{q}_s$ and $q_s$ being the self-overlap among replicas. The so-called entropic potential $\Psi_{s}$ is a function of the input covariance $\Sigma_{\setminus i}$:
\begin{multline}
        \Psi_s = \underset{L\rightarrow \infty}{\mbox{lim}} \frac{\hat{r}^2}{2L} \mbox{tr}\left(\left(\Sigma_{\setminus i}\right)^t\Sigma_{\setminus i}\left( \nu \mathbb{I} + \delta \hat{q} \Sigma_{\setminus i} \right)^{-1} \right)\\
        + \underset{L\rightarrow \infty}{\mbox{lim}} \frac{\hat{q}}{2L}  \mbox{tr}\left(\Sigma_{\setminus i}\left( \nu \mathbb{I} + \delta \hat{q} \Sigma_{\setminus i}\right)^{-1}\right),
\end{multline}
while the energetic potential $\Psi_e$ only depends on the specific choice of the loss function. As shown in the SM, for the optimization problem in eq.\eqref{eq:simplified_optimization} $\Psi_e$ is given by: 
\begin{equation}
    \begin{split}        
        \Psi_e &= -\frac{1}{2 \left( 1+ \delta q \right)} \left( \frac{1}{\nu} + \rho + q - 2r \right).
    \end{split}
\end{equation}

Using this approach, we estimated analytically the generalization loss in Eq. \ref{eq: generalization_loss_replica} as a function of the rescaled number 
of samples $\alpha= M/L$ (see supplementary materials for details).
The result is shown in the right panel of \cref{fig:peak_factored}. As can be noticed, the 
test loss \emph{increases} in the small 
data regime, before peaking at
$\alpha = 1$. This value corresponds to the interpolation threshold, which is the largest number of samples that the neural
network can perfectly fit, which in fact happens at $M=L$. Below this threshold, the model overfits to its
training data; beyond this threshold, the generalization error decreases
monotonically with the training set size; for large $\alpha$, we found
$\epsilon_g \sim \alpha^{-1/2}$. A similar peak in the generalization loss has
been observed in supervised learning~\cite{opper1996statistical} and it is
connected to the well-known ``double descent'' curve observed in deep neural
networks in the presence of label noise~\cite{nakkiran2021deep}. There, the peak is a consequence of overfitting induced by the noise in the labels and it appears after an initial decay of the test loss at small $\alpha$. In
the self-supervised learning regime explored in this work, we find instead that the peak appears naturally as a consequence of the
intrinsic stochasticity of the inputs. Indeed, in masked language modeling, the labels are a part of the input itself. The noise affecting the labels is thus highly correlated to that affecting the input. If the noise in the input is too high, the model starts immediately overfitting and the initial descent is not observed. The absence of the initial descent can be therefore ascribed to the high-level of the noise in the input.

We verify the predictions of the replica theory by plotting the generalization loss of a single layer of
factored self-attention trained on the generalized Potts Model in the setting of \cref{fig:factored_vs_vanilla} at small
regularisation (right side of \cref{fig:peak_factored}). We see the same qualitative
behaviour as predicted by replica theory, even though in this case we did not
apply any of the assumptions required for the replica analysis (the mean field
limit and the usage of a full-rank $J$ matrix and of a $U$ matrix fixed to the
identity). In particular, the test loss increases when adding more data for
small~$\alpha$. The only difference between the plots is the location of the
peak. For the square loss that we analyzed with replicas, as we already commented the peak is at the
interpolation threshold $M=L$. For the simulations with logistic loss, the peak appears at the linearly separability threshold, which is the largest number of points a linear classifier can classify correctly, and which can be larger than one~\cite{gardner1989three,gerace2020generalisation}.

\paragraph*{Concluding perspectives.}

In this work, we have characterised the probability distributions that a single layer of self-attention can learn when trained on a masked language modeling task, considered as a simple prototype of self-supervised learning. In particular, we have shown analytically and numerically that with a single factored-attention layer, it is possible to exactly reconstruct the couplings of a generalized Potts model with two-body
interactions between both sites and colors. More precisely, we showed that training factored
self-attention  on the MLM objective is equivalent to solving the inverse
Potts problem using the pseudo-likelihood method~\cite{besag1975statistical,
  cocco2011adaptive, ricci2012bethe, ekeberg2013improved, cocco2018inverse}, and
therefore it yields consistent estimators of the parameters. These findings make
factored attention a powerful, theoretically-driven building block for deep
transformers. Our replica analysis of
self-attention enabled us to compute the generalization loss of the model
exactly and yielded a non-trivial generalization behaviour. 

Learning higher-order interactions will require additional
layers: a detailed study of how this can be achieved is an interesting direction for future research. It will be
interesting to also study the learning \emph{dynamics} of self-attention
using methods from statistical
physics~\cite{saad1995exact, biehl1995, goldt2020dynamics, veiga2022phase}, both
on MLM and on supervised tasks~\cite{zhang2021pointer, zhang2022unveiling,
  abbe2022learning, seif2022impact}.  In short, our work clarifies the limits  of standard self-attention trained on data where two-body
interactions dominate and highlights the potential
of factored attention as a component of transformer models.

\begin{acknowledgments}
  We thank Marc Mézard, Manfred Opper, and Lenka Zdeborová for stimulating discussions, and Santiago Acevedo for critically reading the manuscript. 
\end{acknowledgments}

\bibliography{refs}
\clearpage

\appendix

\section*{Supplemental material}
\section{Numerical details}

The numerical simulations were perfomed using JAX \cite{jax2018github}. Both the {\it factored} attention layer and the vanilla transformer architecture were optimised using SGD with mini-batch size of 100 and a cosine annealear as the learning rate decay scheduler, both standard choices in the literature. The initial learning rate was adjusted to the specific simulations, choosing it between 0.1 and 0.01. The vanilla transformer code has been taken from Ref.~\cite{lippe2022uvadlc}, with no modifications made. In particular, as already pointed out in the main text, each element $\mathbf{s}_i$ of a sequence $\mathbf{s} = \left( \mathbf{s}_1, ..., \mathbf{s}_L\right)$ is first transformed into a token $\mathbf{x}_i = \mathbf{e}_i + \mathbf{p}_i$, with  $\mathbf{e}_i$ being the embedding of $\mathbf{s}_i$ and $\mathbf{p}_i$ being the positional encoding. The tokenized sequences are then fed to a layer made of two distinct sub-layers. The first sub-layer is composed of a single-head attention, while the second sub-layer contains a two layer fully connected neural network. The inputs of both sublayers are connected to their outputs through skip-connections, and layer normalization is then applied. 

Finally, there is an output layer consisting of a linear transformation from the $d$- to the $C$-dimensional space, in order to obtain a probability distribution over the colors through the {\it softmax} non-linearity. For a graphic visualization of the transformer encoder architecture, reference can be made to the original paper of \citet{vaswani2017attention}. Below is the list of transformer hyperparameters used for the simulations of fig.~\ref{fig:factored_vs_vanilla}: embedding dimension 20, number of heads 1, number of layers 1-3, dropout probability 0.0, number of classes 20.

The dataset was generated using Gibbs sampling, starting from a random sequence of $L=20$ sites and $C=20$ Potts colors and cyclically sampling the spins by exploiting the knowledge of the exact conditional probabilities, eq.~\eqref{eq:exact_marginals}. In order to decorrelate the samples, 10000 Gibbs sweeps were made between each of the two saved configurations.

The simulations on the left panel of fig. \ref{fig:peak_factored}, have been performed by sampling the input data-points from a multivariate Gaussian distribution and the masked token from the same distribution, conditioned on the other elements in the sequence. The optimization problem in eq. \eqref{eq:simplified_optimization} is then solved in closed-form thanks to the Moore-Penrose inverse as in Ref.~\cite{gerace2020generalisation}.

\section{Replica analysis of self-attention}%
\label{sec:supp:replica}

In this appendix we discuss in detail the replica analysis of a factored single layer self-attention. The computation builds upon the recent advances in statistical physics of learning, concerning the extension of replica theory to structured data \cite{gerace2020generalisation,goldt2022gaussian,loureiro2021learning}. In the following we will show how to slightly modify this new approach in order to deal with masked language modeling tasks, under the following simplified assumptions: mean-field limit, full-rank J matrix, U matrix fixed to the identity and thus not learned.

\subsection{Statistical physics formulation of machine learning problems}\label{sec: BG-formulation}

Statistical physics considers learning as a dynamical and exploratory process across the space of the learnable parameters. At equilibrium, these parameters are assumed to follow a Boltzmann-Gibbs distribution, where the role of the Hamiltonian is actually played by the loss function:
\begin{equation}
\begin{split}
    \pi_{\beta}\left( \mathbf{A}_i, \mathcal{D} \right) &= \frac{P\left( \mathbf{A}_i \right)}{\mathcal{Z}_\beta} e^{-\beta \sum_{\mu = 1}^M \ell\left( m^{\mu}_i , \frac{\mathbf{A}_i \cdot \mathbf{m}^{\mu}_{\setminus i}}{\sqrt{L}}\right)} \\
    &= \frac{P_A\left( \mathbf{A}_i \right)}{\mathcal{Z}_\beta} \prod_{\mu = 1}^M P_G \left( m^{\mu}_i| \frac{\mathbf{A}_i \cdot \mathbf{m}^{\mu}_{\setminus i}}{\sqrt{L}}\right) 
    \label{eq:bg}
\end{split}
\end{equation}
with $\beta$ being the inverse temperature and $\mathcal{D}$ the training set. In the zero temperature limit (i.e. $\beta \rightarrow \infty$), the Boltzmann-Gibbs distribution concentrates around the minima of the loss function, which are merely the solutions of the optimization problem in eq. \ref{eq:simplified_optimization}:
\begin{equation}
    \hat{\mathbf{A}}_i \underset{\beta \rightarrow \infty}{=} \mathbb{E}_{\pi_\beta} \left[ \mathbf{A}_i \right] 
\end{equation}
Up to this point, re-framing a machine learning problem in terms of statistical physics did not seem to be very advantageous since sampling from a high-dimensional Boltzmann-Gibbs distribution is known to be impracticable. 

This is where the replica theory comes into play. In particular, it states that, in the high-dimensional limit (i.e. $M,L \rightarrow \infty$ with $\alpha \equiv M/L \sim O\left( 1\right)$) the free-energy of a learning system concentrates around its typical value over the input data distribution:
\begin{equation}
     f_{\beta} = -\underset{L \rightarrow \infty}{\mbox{lim}} \frac{1}{L}  \mathbb{E}_{\{ \mathbf{m}^{\mu}_{\setminus i}, m^{\mu}_i\}} \left[ \mbox{log} \mathcal{Z}_\beta\right].
     \label{eq: typical_free_energy}
\end{equation}
As we will see in the next section, this expectation can be tackled by means of the replica trick. From this quantity, all the high-dimensional metrics of interests, can be computed as a function of simple scalar quantities. This is for instance the case of the generalization loss in eq. \eqref{eq: generalization_loss_replica}. In particular, the overlap parameters $m_{\star}$ and $q_{\star}$ correspond to practically measurable quantities over different realisation of the training set, involving the estimator of the $i$th row of the interaction matrix:  
\begin{equation}
\label{eq:supp-order-parameters}
 \begin{split}
     q^{\star} = \frac{\hat{\mathbf{A}}^t_i \Sigma_{\setminus i} \hat{\mathbf{A}}_i}{L}   \hspace{5mm} r^{\star} = - \frac{\mathbf{J}_i^t \Sigma_{\setminus i} \hat{\mathbf{A}}_i}{\nu L}.
 \end{split}
 \end{equation}
In the next section we will outline the main steps of the replica trick leading to the generalization loss formula in eq. \eqref{eq: generalization_loss_replica}.

\subsection{Replica Calculation}

As anticipated in the previous section, the replica trick allows to compute the typical value of the free-energy density in eq. \eqref{eq: typical_free_energy} by expressing this quantity as a function of the solely replicated partition function $\mathcal{Z}_\beta^n$, obtained by constructing $n > 0$ different and independent copies of the same learning system:
\begin{equation}
     f_{\beta} = -\underset{n \rightarrow 0^+}{\mbox{lim}}\frac{d}{dn}\underset{L \rightarrow \infty}{\mbox{lim}} \left[ \frac{1}{L}  \mathbb{E}_{\{ \mathbf{m}^{\mu}_{\setminus i}, m^{\mu}_i\}} \mathcal{Z}_\beta^n \right].
\end{equation}

\paragraph{Average over the training set.} As a first step, the replica calculation focuses on the expectation of the replicated partition function over the training set, which, written in a more explicit form, looks like:
\begin{equation}
\begin{split}
\mathbb{E}_{\{ \mathbf{m}^{\mu}_{\setminus i}, m^{\mu}_i\}} \mathcal{Z}^n_{\beta} &= \int \prod_{a = 1}^n d\mathbf{A}_i^a \prod_{a = 1}^n \ P_A\left( \mathbf{A}_i^a\right)\\ 
&\times \prod_{\mu = 1}^M \mathbb{E}_{ m^{\mu}_i| \mathbf{m}^{\mu}_{\setminus i}} \mathbb{E}_{\mathbf{m}^{\mu}_{\setminus i}}\left[ P_G\left( m_i^{\mu}\mid \frac{\mathbf{A}^a_i \cdot \mathbf{m}^{\mu}_{\setminus i}}{\sqrt{L}}\right)\right]
\label{eq: averaged_replicated_partition_function},
\end{split}
\end{equation}
with $P_G$ and $P_A$ being respectively the Gibbs and the Gaussian measure associated with the $i$th row of the attention matrix as in eq. \eqref{eq:bg}. Indeed, as already pointed out in the main manuscript, the interaction matrix is drawn from the Gaussian Orthogonal Ensemble, therefore its rows will correspond to Gaussian random vectors. 
At this point, we can notice an important aspect of MLM tasks. In this case, the labels are not provided by a teacher vector as in standard teacher-student settings. On the contrary, the masked tokens are directly sampled from the input distribution by conditioning over all the other elements composing the sequence:
\begin{equation}
\begin{split}
    m_i^{\mu} &\sim \frac{P_J\left( \mathbf{J}_i\right)}{\sqrt{2\pi \nu^{-1}}}e^{-\frac{1}{2\nu^{-1}} \left( m^{\mu}_i + \frac{\mathbf{J}_i \cdot \mathbf{m}^{\mu}_{\setminus i}}{\nu\sqrt{L}} \right)^2}\\ 
    &= P_J\left( \mathbf{J}_i\right) P_0 \left( m_i^{\mu}\mid \frac{\mathbf{J}_i \cdot \mathbf{m}^{\mu}_{\setminus i}}{\nu\sqrt{L}} \right).
\end{split}
\end{equation}
Note that, the noise in the labels arises a consequence of the one already affecting the input data points, meaning that its intensity can not be chosen independently from the intrinsic stochasticity of the input. Due to these considerations, by explicitly expressing the outer expectation in eq. \eqref{eq: averaged_replicated_partition_function}, we then obtain 
\begin{equation}
\begin{split}
\mathbb{E}_{\{ \mathbf{m}^{\mu}_{\setminus i}, m^{\mu}_i\}} \mathcal{Z}^n_{\beta} &= \int d\mathbf{J}_i \ P_J\left( \mathbf{J}_i\right) \prod_{a = 1}^n \left(\int d\mathbf{A}_i^a \ P_A\left( \mathbf{A}_i^a\right)\right) \times\\
&\times  \prod_{\mu =1}^M \left( \int dm_i^{\mu}\mathbb{E}_{\mathbf{m}^{\mu}_{\setminus i}} \left[ P_0 \left( m_i^{\mu}\mid \frac{\mathbf{J}_i \cdot \mathbf{m}^{\mu}_{\setminus i}}{\nu\sqrt{L}} \right) \times\right.\right.\\
&\left. \left. \times P_G \left( m_i^{\mu}\mid \frac{\mathbf{A}^{a}_i \cdot \mathbf{m}^{\mu}_{\setminus i}}{\sqrt{L}} \right)\right]\right)
\label{eq: averaged_replicated_partition_function_after_label_expectation}.
\end{split}
\end{equation}
To compute the expectation over the input sequences with a masked element at position $i$, we first define the pre-activations as:
\begin{equation}
    h^{\mu}_a = \frac{\mathbf{A}^a_i \cdot \mathbf{m}^{\mu}_{\setminus i}}{\sqrt{L}} \hspace{10mm} z^{\mu} = -\frac{\mathbf{J}_i \cdot \mathbf{m}^{\mu}_{\setminus i}}{\nu\sqrt{L}},
\end{equation}
then we express these definitions in terms of Dirac-deltas and their corresponding integral representation:
\begin{equation}
    \begin{split}
        & 1 \propto \int \prod_{\mu = 1}^M \prod_{a=1}^n \frac{dh_a^{\mu} d\hat{h}^{\mu}_a}{2\pi} \prod_{\mu = 1}^M e^{i \hat{h}^{\mu}_a \left(  h_a^{\mu} - \frac{\mathbf{A}^a_i \cdot \mathbf{m}^{\mu}_{\setminus i}}{\sqrt{L}} \right)}\\
        & 1 \propto  \int \prod_{\mu = 1}^M \frac{dz^{\mu} d\hat{z}^{\mu}}{2\pi} \prod_{\mu = 1}^M \ e^{i \hat{z}^{\mu} \left(  z^{\mu} + \frac{\mathbf{J}_i \cdot \mathbf{m}^{\mu}_{\setminus i}}{\nu\sqrt{L}} \right)}. 
    \end{split}
\end{equation}
By plugging these factors one into eq. \eqref{eq: averaged_replicated_partition_function_after_label_expectation}, we then get:
\begin{equation}
\begin{split}
&\mathbb{E}_{\{ \mathbf{m}^{\mu}_{\setminus i}, m^{\mu}_i\}} \mathcal{Z}^n_{\beta} = \int d\mathbf{J}_i \ P_J\left( \mathbf{J}_i\right) \int \prod_{a = 1}^n d\mathbf{A}_i^a \ \prod_{a = 1}^n P_A\left( \mathbf{A}_i^a\right) \times\\
&\times \int \prod_{\mu=1}^M dm_i^{\mu} \int \prod_{\mu=1}^M \frac{dz^{\mu} d\hat{z}^{\mu}}{2\pi} \prod_{\mu = 1}^M e^{i \hat{z}^{\mu} z^{\mu}} \prod_{\mu = 1}^M P_0 \left( m_i^{\mu}\mid z^{\mu} \right)\times\\
&\times \int \prod_{\mu=1}^M \prod_{a=1}^n \frac{dh_a^{\mu} d\hat{h}_a^{\mu}}{2\pi} \prod_{\mu = 1}^M \prod_{a=1}^n e^{i \hat{h}_a^{\mu} h_a^{\mu}} \prod_{\mu = 1}^M \prod_{a=1}^n P_G \left( m_i^{\mu}\mid h_a^{\mu} \right)\times\\
&\times \prod_{\mu=1}^n \mathbb{E}_{\mathbf{m}^{\mu}_{\setminus i}} \left[ e^{i \hat{z}^{\mu} \frac{\mathbf{J}_i \cdot \mathbf{m}^{\mu}_{\setminus i}}{\nu\sqrt{L}}} \prod_{a=1}^n e^{- i \hat{h}_a^{\mu} \frac{\mathbf{A}^{a}_i \cdot \mathbf{m}^{\mu}_{\setminus i}}{\sqrt{L}}}\right] 
\label{eq: averaged_replicated_partition_function_after_dirac_deltas}.
\end{split}
\end{equation}
The expectation over the masked input sequences is a simple multivariate Gaussian integral, whose solution is given by:
\begin{equation}
    \begin{split}
        &\mathbb{E}_{\mathbf{m}^{\mu}_{\setminus i}} \left[ e^{i \hat{z}^{\mu} \frac{\mathbf{J}_i \cdot \mathbf{m}^{\mu}_{\setminus i}}{\nu\sqrt{L}}} \prod_{a=1}^n e^{- i \hat{h}_a^{\mu} \frac{\mathbf{A}^{a}_i \cdot \mathbf{m}^{\mu}_{\setminus i}}{\sqrt{L}}}\right] =
        e^{-\frac{1}{2} \frac{\mathbf{J}_i^t \Sigma_{\setminus i} \mathbf{J}_i}{\nu^2 L} \left(z^{\mu}\right)^2}\times \\
        & \times e^{\sum_{a=1}^n \frac{\mathbf{J}_i^t \Sigma_{\setminus i} \mathbf{A}_i^{a}}{\nu L} \hat{h}^{\mu}_a \hat{z}^{\mu} -\frac{1}{2} \sum_{a,b = 1}^n  \frac{\left(\mathbf{A}_i^{a} \right)^t \Sigma_{\setminus i} \mathbf{A}_i^{a}}{L} \hat{h}^{\mu}_a \hat{h}^{\mu}_b}
    \end{split}
\end{equation}
where, as already pointed out in the main text, $\Sigma_{\setminus i}$ corresponds to the covariance matrix of the masked input sequences, that is the input covariance matrix without the contribution of the row and column associated to the $i$th masked token. By replacing the solution of the expectation in eq. \eqref{eq: averaged_replicated_partition_function_after_dirac_deltas}, we then get: 
\begin{equation}
\begin{split}
&\mathbb{E}_{\{ \mathbf{m}^{\mu}_{\setminus i}, m^{\mu}_i\}} \mathcal{Z}^n_{\beta} = \int d\mathbf{J}_i \ P_J\left( \mathbf{J}_i\right) \int \prod_{a = 1}^n d\mathbf{A}_i^a \ \prod_{a = 1}^n P_A\left( \mathbf{A}_i^a\right) \times\\
&\times \int \prod_{\mu=1}^M dm_i^{\mu} \int \prod_{\mu=1}^M \frac{dz^{\mu} d\hat{z}^{\mu}}{2\pi} \prod_{\mu = 1}^M e^{i \hat{z}^{\mu} z^{\mu}} \prod_{\mu = 1}^M P_0 \left( m_i^{\mu}\mid z^{\mu} \right)\times\\
&\times \int \prod_{\mu=1}^M \prod_{a=1}^n \frac{dh_a^{\mu} d\hat{h}_a^{\mu}}{2\pi} \prod_{\mu = 1}^M \prod_{a=1}^n e^{i \hat{h}_a^{\mu} h_a^{\mu}} \prod_{\mu = 1}^M \prod_{a=1}^n P_G \left( m_i^{\mu}\mid h_a^{\mu} \right)\times\\
&\times \prod_{\mu=1}^n e^{-\frac{1}{2} \frac{\mathbf{J}_i^t \Sigma_{\setminus i} \mathbf{J}_i}{\nu^2 L} \left(z^{\mu}\right)^2 + \sum_{a=1}^n \frac{\mathbf{J}_i^t \Sigma_{\setminus i} \mathbf{A}_i^{a}}{\nu L} \hat{h}^{\mu}_a \hat{z}^{\mu} -\frac{1}{2} \sum_{a,b = 1}^n  \frac{\left(\mathbf{A}_i^{a} \right)^t \Sigma_{\setminus i} \mathbf{A}_i^{a}}{L} \hat{h}^{\mu}_a \hat{h}^{\mu}_b}
\label{eq: averaged_replicated_partition_function_after_x_average}.
\end{split}
\end{equation}

\paragraph{Rewriting the averaged replicated partition function in terms of saddle-point integrals.} As a consequence of the average over the training set, the different replicas are now interacting among each other through the following set of overlap parameters:
\begin{equation}
    \rho = \frac{\mathbf{J}_i^t \Sigma_{\setminus i} \mathbf{J}_i}{\nu^2 L} \hspace{5mm} r_a = -\frac{\left(\mathbf{A}_i^{a}\right)^t \Sigma_{\setminus i} \mathbf{J}_i}{\nu L} \hspace{5mm} q_{ab} = -\frac{\left(\mathbf{A}_i^{a}\right)^t \Sigma_{\setminus i} \mathbf{A}_i^{b}}{L}. 
\end{equation}
Once again, to proceed further in the calculation, we can insert their definition by means of Dirac-deltas and their integral representation:
\begin{equation}
    \begin{split}
        & 1 \propto \int \frac{d\rho d\hat{\rho}}{2\pi} \ e^{i \hat{\rho} \left( \nu^2 L\rho - \mathbf{J}_i^t \Sigma_{\setminus i} \mathbf{J}_i \right)}\\
        & 1 \propto \int \prod_{a = 1}^n \frac{dr_a d\hat{r}_a}{2\pi} \prod_{a = 1}^n \ e^{i \hat{r}_a \left( -\nu L r_a - \left(\mathbf{A}^a_i\right)^t \Sigma_{\setminus i} \mathbf{J}_i  \right)}\\
        & 1 \propto \int \prod_{a \le b} \frac{dq_{ab} d\hat{q}_{ab}}{2\pi} \prod_{a \le b} \ e^{i \hat{q}_{ab} \left( q_{ab} - \left(\mathbf{A}^a_i\right)^t \Sigma_{\setminus i}\mathbf{A}^{b}_i  \right)}.
    \end{split}
\end{equation}
By substituting the overlap definition in eq. \eqref{eq: averaged_replicated_partition_function_after_x_average}, plugging in the corresponding factors one and performing the change of variables: $i\hat{\rho} \rightarrow -\hat{\rho}$, $i\hat{r}_a \rightarrow \hat{r}_a$ and $i\hat{q}_{ab} \rightarrow \hat{q}_{ab}$, we can rewrite the averaged replicated partition function in terms of saddle-point integrals over the overlap parameters:
\begin{equation}
\mathbb{E}_{\{ \mathbf{m}^{\mu}_{\setminus i}, m^{\mu}_i\}} \mathcal{Z}^n_{\beta} = \int \frac{d\rho d\hat{\rho}}{2\pi} \int \prod_{a=1}^n \frac{d r_a d\hat{r}_a}{2\pi} \int \prod_{a \leq b} \frac{dq_{ab} d\hat{q}_{ab}}{2\pi} \ e^{L \Psi^{\left( n \right)} }\label{eq: averaged_replicated_partition_function_saddle}
\end{equation}
where the action $ \Psi^{\left( n \right)}$ is a non trivial function of the overlap parameters:
\begin{equation}
   \Psi^{\left( n \right)} = -\nu^2 \rho \hat{\rho} + \nu \sum_{a=1}^n r_a\hat{r}_a - \sum_{a \leq b} q_{ab} \hat{q}_{ab} + \frac{1}{L}\Psi_s + \frac{M}{L} \Psi_e 
   \label{eq: action}
\end{equation}
where $\Psi_s$ and $\Psi_e$ are the so-called entropic and energetic potential and, in the specific case of a single-layer factored attention, are given by:
\begin{equation}
\begin{split}
    \Psi_s &= \mbox{log} \left[ \int d\mathbf{J}_i P_J\left( \mathbf{J}_i  \right) \int \prod_{a=1}^n \left(d\mathbf{A}_i^{a} P_A\left( \mathbf{A}_i^{a}  \right)\right) \times \right.\\
    &\left. \times e^{\hat{\rho} \mathbf{J}_i^t \Sigma_{\setminus i} \mathbf{J}_i + \sum_{a=1}^n \hat{r}_a \left( \mathbf{A}_i^{a} \right)^t \Sigma_{\setminus i} \mathbf{J}_i + \sum_{a \leq b} \hat{q}_{ab} \left( \mathbf{A}_i^{a} \right)^t \Sigma_{\setminus i} \mathbf{A}_i^{b} }\right]\\
    \Psi_e &=  \mbox{log} \left[ \int dm_i \int \frac{dzd\hat{z}}{2\pi} \ e^{-\frac{\rho}{2} \hat{z}^2 + i \hat{z}z} P_0\left( m_i | z \right) \right. \times \\
    &\left. \times \int \prod_{a=1}^n \frac{dh_a d\hat{h}_a}{2\pi} \ e^{-\frac{1}{2} \sum_{a,b=1}^n q_{ab} \hat{h}_a \hat{h}_b - \hat{z} \sum_{a=1}^n r_a \hat{h}_a} \right. \times\\
    & \times \left. e^{i \sum_{a=1}^n\hat{h}_a h_a}P_G\left( m_i | h_a \right)\right].
    \label{eq: entropic_and_energetic_potential}
\end{split}
\end{equation}
Note that, we have drop the dependency of $\Psi_e$ on the $\mu$ index since all the $\mu$-dependent terms decouple with respect to $\mu$. 

\paragraph{Replica Symmetric Assumption.} To proceed further in the calculation, we need to assume a specific replica structure. Since all replicas have been introduced independently from each other with no specific differences among them, it seems natural to assume that replicas should all play the same role and that, therefore, the overlap parameters should not depend on the specific replica index. In particular, under the replica symmetric ansatz, we assume:
\begin{equation}
\begin{split}
    q_{ab}&=
    \begin{cases}
      g & \text{if}\ a = b \\
      q & \text{otherwise}
    \end{cases} \hspace{5mm} -i\hat{q}_{ab}=
    \begin{cases}
      -\frac{\hat{g}}{2} & \text{if}\ a = b \\
      \hat{q} & \text{otherwise}
    \end{cases}\\\\
r_a &= r \hspace{2mm}  \hspace{24mm} -i\hat{r}_a = \hat{r} \hspace{5mm} \forall a\\ 
\end{split}
\end{equation}
By plugging the replica symmetric assumption in eq. \eqref{eq: action}-\eqref{eq: entropic_and_energetic_potential} and applying the following Hubbard Stratonovich transformations:
\begin{equation}
\begin{split}
e^{\frac{\hat{q}}{2} \sum_{a\leq b} \left( \mathbf{A}^a \right)^t \Sigma_{\setminus i} \mathbf{A}^b} &= \int \mathcal{D}\boldsymbol{\xi} \ \mbox{exp}\left(\sum_{a = 1}^n\left( \mathbf{A}_i^a \right)^t\left( \hat{q}\Sigma_{\setminus i} \right)^{1/2} \boldsymbol{\xi}\right)\\
e^{-\frac{q}{2} \sum_{a,b = 1}^n \hat{h}_a \hat{h}_b} &= \int \mathcal{D}\xi \ \mbox{exp}\left( i \sqrt{q} \sum_{a=1}^n\hat{h}^2_a \ \xi \right),
\end{split}
\end{equation}
with $\xi \sim \mathcal{N}\left( 0, 1\right)$, we then get the expression for the replica symmetric action:
\begin{equation}
\begin{split}
   \Psi^{\left( n \right)} &= -\nu^2 \rho \hat{\rho} + \nu r\hat{r} + \frac{n}{2} \left( \delta q+q \right)\left(\delta \hat{q} - \hat{q}\right) - \frac{n\left( n-1\right)}{2} q \hat{q}\\
   &+ \frac{1}{L}\Psi_s + \frac{M}{L} \Psi_e 
   \label{eq: action_after_RS}
\end{split}
\end{equation}
where we have defined $\delta\hat{q} = \hat{g} + \hat{q}$ and $\delta q = g-q$.  the replica symmetric potentials $\Psi_s$ and $\Psi_e$ are are given by:
\begin{equation}
    \begin{split}
     \Psi_s &= \mbox{log} \left[ \int d\mathbf{J}_i P_J\left( \mathbf{J}_i  \right) e^{\hat{\rho} \mathbf{J}_i^t \Sigma_{\setminus i} \mathbf{J}_i}
     \int \mathcal{D}\boldsymbol{\xi} \times\right.\\
     &\left.\times\left(\int d\mathbf{A}_i P_A\left( \mathbf{A}_i  \right) e^{ \hat{r} \mathbf{A}_i^t \Sigma_{\setminus i} \mathbf{J}_i  -\frac{\hat{l}}{2} \mathbf{A}_i^t \Sigma_{\setminus i} \mathbf{A}_i + \mathbf{A}^t_i \left( \hat{q} \Sigma_{\setminus i} \right)^{1/2} \boldsymbol{\xi} } \right)^n\right]\\
    \Psi_e &=  \mbox{log} \left[ \int dm_i \int \frac{dzd\hat{z}}{2\pi} \ e^{-\frac{\rho}{2} \hat{z}^2 + i \hat{z}z} P_0\left( m_i | z \right) \int \mathcal{D}\xi \times \right.\\ 
    & \times \left.\left( \int\frac{dh d\hat{h}}{2\pi} \ e^{-\frac{l}{2} \hat{h}^2 + i \sqrt{q}\xi \hat{h} - m\hat{z}\hat{h} + i \hat{h}h} P_G\left( m_i | h \right)\right)^n\right]
    \end{split}
    \label{eq: entropic_and_energetic_potential_after_RS}
\end{equation}

\paragraph{Zero Replica limit.} By taking the limit of $n\rightarrow 0 $ in eq. \eqref{eq: action_after_RS}-\eqref{eq: entropic_and_energetic_potential_after_RS} and solving the integrals with respect to the $\hat{z}$ and $\hat{h}$ variables, we then get the following expression for the action potential: 
\begin{equation}
    \Psi^{\left( n \rightarrow 0 \right)} = \nu r \hat{r} + \frac{1}{2} \left( l+q \right)\left(\hat{l} - \hat{q}\right) + \frac{1}{2} q \hat{q} + \frac{1}{L}\Psi_s^{\left( n \rightarrow 0 \right)} + \frac{M}{L}\Psi_e^{\left( n \rightarrow 0 \right)} 
\end{equation}
with the entropic and energetic potentials in the zero replicas limit given by:
\begin{equation}
\begin{split}
    &\Psi_s^{\left( n \rightarrow 0 \right)} = \int \mathcal{D}\boldsymbol{\xi} \int d\mathbf{J}_i P_J\left( \mathbf{J}_i  \right) \times\\
    & \mbox{log}\left[\int d\mathbf{A}_i P_A\left( \mathbf{A}_i  \right) e^{ \hat{r} \mathbf{A}_i^t \Sigma_{\setminus i} \mathbf{J}_i -\frac{\hat{l}}{2} \mathbf{A}_i^t \Sigma_{\setminus i} \mathbf{A}_i + \mathbf{A}^t_i \left( \hat{q} \Sigma_{\setminus i} \right)^{1/2} \boldsymbol{\xi} } \right]\\
    &\Psi_e^{\left( n \rightarrow 0 \right)} =  \int \mathcal{D}\xi\int dm_i \int \frac{dz}{\sqrt{2\pi \left( \rho - \frac{r^2}{q}\right)}} \ e^{-\frac{\left(z + \frac{r}{\sqrt{q}}\xi\right)^2}{2\left(\rho - \frac{r^2}{q} \right)}}\times\\    
    & P_0\left( m_i | z \right) \mbox{log} \left[\int\frac{dh}{\sqrt{2\pi l}} \ e^{- \frac{\left( h + \sqrt{q}\xi\right)^2}{2l}} P_G\left(m_i|h \right)\right] 
    \label{eq: averaged_replicated_partition_function_zero_replicas}
\end{split}    
\end{equation}
Note that, as in standard teacher-student settings \cite{gerace2020generalisation}, in order to avoid divergent terms in this limit, the overlap $\rho$ and its conjugate $\hat{\rho}$ need to be constrained to $\mathbb{E}_{\mathbf{J}_i}\left[ \mathbf{J}_i^t \Sigma_{\setminus i} \mathbf{J}_i \right]/\nu^2 $ and $0$ respectively.  

\paragraph{Typical Free-Energy density.} Having determined the expression for the replicated partition function in the zero-temperature limit, we can actually compute the typical free-energy density as:

\begin{equation}
\begin{split}
     f_{\beta} &= -\underset{n \rightarrow 0^+}{\mbox{lim}}\frac{d}{dn}\underset{L \rightarrow \infty}{\mbox{lim}} \left[ \frac{1}{L}  \mathbb{E}_{\{ \mathbf{m}^{\mu}_{\setminus i}, m^{\mu}_i\}} \mathcal{Z}_\beta^n \right]\\ 
     &= \underset{L \rightarrow \infty}{\mbox{lim}} \frac{1}{L} \int \frac{d\rho d\hat{\rho}}{2\pi} \int \frac{d r d\hat{r}}{2\pi} \int \frac{dq d\hat{q}}{2\pi} \ e^{L \Psi^{\left( n \rightarrow 0 \right)} } .
\end{split}
\end{equation}

In the high-dimensional limit, we can solve the integrals over the overlap parameters by saddle-point, thus getting: 

\begin{equation}
\begin{split}
     f_{\beta} &= \underset{q,r,\delta q, \hat{q}, \hat{r}, \delta \hat{q}}{\mbox{extr}} \left[\nu r \hat{r} + \frac{1}{2} \left( \delta q+q \right)\left(\delta \hat{q} - \hat{q}\right) + \frac{1}{2} q \hat{q}\right.+\\
     &\left. + \underset{L \rightarrow \infty}{\mbox{lim}} \frac{1}{L}\Psi_s^{\left( n \rightarrow 0 \right)} + \alpha \Psi_e^{\left( n \rightarrow 0 \right)}\right],
\end{split}
\end{equation}
where the values of the overlap parameters extremizing the typical free-energy density can therefore be determined by solving the following system of coupled saddle-point equations:
\begin{equation}
\begin{split}
    \begin{cases}
     q=-2\frac{\partial \Psi_s^{\left(n \rightarrow 0\right)}}{ \partial \delta\hat{q}}\\
     \delta q= 2\frac{\partial \Psi_s^{\left(n \rightarrow 0\right)}}{ \partial \hat{q}}\\
     r = - \frac{1}{\nu}\frac{\partial \Psi_s^{\left(n \rightarrow 0\right)}}{ \partial \hat{r}}
    \end{cases} \hspace{10mm}
    \begin{cases}
    \hat{q}= 2 \frac{\partial \Psi_e^{\left(n \rightarrow 0\right)}}{\partial \delta q}\\
    \delta\hat{q} = -2 \frac{\partial \Psi_e^{\left(n \rightarrow 0\right)}}{\partial q}\\
    \hat{r} = -\frac{1}{\nu} \frac{\partial \Psi_e^{\left(n \rightarrow 0\right)}}{\partial r} 
    \end{cases}\\\\
\end{split}.
\label{eq: saddle-point}
\end{equation}
Up to this point, we have performed the replica calculation in full generality, without specifying neither the interaction matrix nor the loss function. In the next section, we will evaluate the typical free-energy density for the specific MLM task of eq. \eqref{eq:simplified_optimization} under the simplified assumptions of the sec. \textit{The sample complexity of self-attention} of the main text.

\paragraph{Zero temperature limit and Gaussian Priors.} As already pointed out in the main text, the interaction matrix $J$ is sampled from the GOE, it is then natural to assume a Gaussian prior on the $i$th row of the attention matrix:

\begin{equation}
    P_A\left( \mathbf{\mathbf{A}} \right) = \frac{1}{\sqrt{2\pi}} e^{\beta \lambda\mathbf{A}_i^t \mathbf{A}_i}
\end{equation}
with $\beta$ being the inverse temperature parameter, while $\lambda$ is the $L_2$ regularization strength. Moreover, the optimization problem in eq. \eqref{eq:simplified_optimization} optimizes a square loss to solve the corresponding MLM task. This means that, the Gibbs Measure of eq. \eqref{eq:bg} associated with this task is:
\begin{equation}
    P_G\left( m_i|h \right) \propto e^{-\frac{\beta}{2} \left( m_i - h \right)^2}.
\end{equation}
By plugging these two specific forms of both the prior and the Gibbs measure in eq. \eqref{eq: averaged_replicated_partition_function} and taking the zero temperature limit as exemplified in \cite{gerace2020generalisation, goldt2022gaussian}, we get the following expression for the typical free-energy density in the zero-temperature limit: 

\begin{equation}
\begin{split}
     f &\underset{\beta \rightarrow \infty}{=} \underset{q,r,\delta q, \hat{q}, \hat{r}, \delta\hat{q}}{\mbox{extr}} \left[\nu r \hat{r} - \frac{1}{2} \left( \delta q\hat{q} - q\delta\hat{q} \right)+\right.\\ 
     &\left.+ \underset{L \rightarrow \infty}{\mbox{lim}} \frac{1}{L}\Psi_s^{\left( n \rightarrow 0 \right)} + \alpha \Psi_e^{\left( n \rightarrow 0 \right)}\right],
\end{split}
\end{equation}
with the entropic and energetic potential given by:
\begin{equation}
\begin{split}
    \Psi_s^{n \rightarrow 0} &\underset{\beta \rightarrow \infty}{=} \frac{1}{2} \left[ \hat{r}^2 \mbox{tr} \left(\Sigma_{\setminus i}\right)^t\Sigma_{\setminus i} \left( \beta \lambda \mathbb{I} + \delta\hat{q}\Sigma_{\setminus i} \right)^{-1} +\right.\\
    &\left.+\hat{q} \mbox{tr} \Sigma_{\setminus i}\left( \beta \lambda \mathbb{I} + \delta\hat{q}\Sigma_{\setminus i} \right)^{-1}\right]\\
   \Psi_e^{n \rightarrow 0} &\underset{\beta \rightarrow \infty}{=} -\frac{1}{2} \frac{\nu^{-1} + \rho + q - 2r }{1 + \delta q }
\end{split}
\end{equation}
and $\rho = \mbox{tr}\Sigma_{\setminus i}/\left( \nu^2 L \right) $. Note that, this functional form of the typical free-energy density corresponds exactly to the one of supervised learning with noisy label and Gaussian structured data \cite{goldt2022gaussian}. However, we should point out again that, the variance of the noise in the labels, controlled by the shift factor $\nu$, is a direct consequence of the intrinsic noise already affecting the input. Therefore, it can not be tuned independently from it. This is also reflected in the slightly different functional forms of the saddle-point equations in \eqref{eq: saddle-point}, which, in the zero-temperature limit with Gaussian priors and square loss, are given by: 
\begin{equation}
\begin{split}
    \begin{cases}
     q=\mbox{tr} \left[ \left(\hat{q}\Sigma_{\setminus i} + \hat{r}^2 \Sigma_{\setminus i} \left( \Sigma_{\setminus i} \right)^t\right) \Sigma_{\setminus i} \left( \lambda \mathbb{I} + \delta \hat{q} \Sigma_{\setminus i} \right)^{-2}  \right] \\
     \delta q= \mbox{tr} \left[ \left(\lambda \mathbb{I} + \delta \hat{q}\Sigma_{\setminus i}\right)^{-1} \Sigma_{\setminus i} \right]  \\
     r = - \frac{\hat{r}}{\nu}\mbox{tr}\left[ \Sigma_{\setminus i}\left(\Sigma_{\setminus i}\right)^t \left( \lambda\mathbb{I} + \delta \hat{q}\Sigma_{\setminus i} \right)^{-1} \right]\\
    \hat{q}= \frac{\nu^{-1} + \rho + q - 2r}{\left( 1 + \delta q \right)^2}\\
    \delta \hat{q} = \frac{1}{1+\delta q}\\
    \hat{r} = -\frac{1}{\nu \left( 1+\delta q\right)} 
    \end{cases}.
\end{split}
\label{eq: saddle-point_zero_temperature}
\end{equation}

As in the case of supervised learning settings \cite{gerace2020generalisation,goldt2022gaussian}, the solution of this system of coupled saddle-point equations in the zero temperature limit allows us to express the generalization loss as shown in Eq.~\ref{eq:eg} of the main text, with the exception that the noise in the labels is the direct consequence of the intrinsic noise of the inputs. 

\end{document}